\documentclass[twocolumn,showpacs,preprintnumbers,amsmath,amssymb,prl]{revtex4}
\usepackage{graphicx}
\input epsf \unitlength=1mm

\begin{document} 
\preprint{APS/123-QED}
\title{Quantum Monte Carlo based on two-body density functional theory for
fermionic many-body systems: application to $^3$He}

\author{Bal{\'a}zs Het{\'e}nyi}

%\altaffiliation[Corresponding author:]{
%                                               bhetenyi@sissa.it}
\author{L. Brualla}
\author{S. Fantoni}
%\author{K. E. Schmidt}\altaffiliation[Permanent address:]{
%                         Dept. of Physics and Astronomy, 
%			 Arizona State University, Tempe, AZ 85287}

\affiliation{SISSA-International School of Advanced Studies, via
Beirut 2-4, Trieste 34014, Italy}

\begin{abstract}
We construct a quantum Monte Carlo algorithm for interacting fermions
using the two-body density as the fundamental quantity.  The central
idea is mapping the interacting fermionic system onto an auxiliary
system of interacting bosons.  The correction term is approximated
using correlated wave-functions for the interacting system, resulting
in an effective potential that represents the nodal surface.  We
calculate the properties of $^3$He and find good agreement with
experiment and with other theoretical work.  In particular our results
for the total energy agree well with other calculations where the same
approximations were implemented but the standard quantum Monte Carlo
algorithm was used.
\end{abstract}
\maketitle
%\section{Introduction}
\label{sec:intro}

Density-functional theory~\cite{Hohenberg64,Kohn65} (DFT) and quantum
Monte Carlo~\cite{Anderson75,Foulkes01} (QMC) are generally thought of
as two distinct approaches to the problem of interacting fermions.
DFT is based on the Hohenberg-Kohn theorems~\cite{Hohenberg64} (HK)
which state that the energy of an interacting fermion system in an
external field can be written as a functional of the density, and that
minimizing the energy as a functional of the density gives the ground
state energy (HK theorems).  Applications of DFT are usually based on
the Kohn and Sham~\cite{Kohn65} method, where an auxiliary
noninteracting system is invoked.  Minimization is achieved with
respect to the orbitals of the auxiliary system.  QMC also involves
minimization of the energy.  One way of minimizing, is to propagate a
trial wavefunction in imaginary time~\cite{Anderson75}, so that it
asymptotically approaches the ground state.

In this letter we present a QMC method derived from an extension of
DFT where the two-body density ($n^{(2)}$) is the fundamental
quantity~\cite{Ziesche96,Levy01}.  As in standard DFT the energy
functional is universal but unknown, thus approximations schemes are
necessary.  In the spirit of the Kohn-Sham ansatz we invoke an
auxiliary system with identical $n^{(2)}$ as the system of interacting
fermions under investigation, but instead of a non-interacting system,
one of interacting bosons.  As in the Kohn-Sham method, minimization
is not performed with respect to the density, but with respect to the
bosonic wavefunction via QMC~\cite{Anderson75}.  This can be done,
since the two-body density can be written in terms of the bosonic
wavefunction.  In our method the sign-problem does not arise
explicitly, thus fixed-node~\cite{Reynolds82} (FN) or
released-node~\cite{Ceperley80} (RN) techniques are not needed.  In
the auxiliary fields Monte Carlo method~\cite{Baer98} the need for FN
or RN is also circumvented, but the sign-problem still manifests in
the phases of the auxiliary fields.

The correction term necessitated by our ansatz is obtained
approximately using correlated basis functions.  The resulting
approximation consists of a two-body and a three-body potential
(effective nodal surface). The appearance of the three-body potential
(and density) in our energy functional is a result of our
approximation scheme.  In principle our approximate energy functional
can still be written as a functional of $n^{(2)}$, since according to
the HK theorems~\cite{Hohenberg64} the three-body density (as all
other observables) is a functional of $n^{(2)}$.  As far as the method
developed here is concerned, the minimization itself is performed with
respect to the bosonic wavefunction, thus higher-order potentials are
easily handled.

We apply our formalism to calculate the total energy, potential
energy, and the structure factor of $^3$He in a range of densities
close to the equilibrium one ($\rho_0=0.273N/\sigma^{-3}$,
$\sigma=2.556$ \AA).  Our model estimates the density that minimizes
the total energy to be slightly less than the experimental result, as
one would expect from the fact that we are not including back-flow
effects.  The calculated energies are in very good agreement with QMC
results at the same level of approximation, and compare reasonably
well with experiment.

Given a system of interacting particles with potential $w$ (including
two-body and one-body potentials) and with two-body density $n^{(2)}$
(the diagonal elements of the two-body density matrix) HK can be
extended as follows:
\begin{itemize}
\item There is a one to one correspondence between $w$ and $n^{(2)}$.
\item The ground state energy of the system can be obtained by
minimizing $E$ as a function of $n^{(2)}$.
\end{itemize}
The proof of these statements is an easy extension of the original
work~\cite{Hohenberg64}, and can be extended to $N$-representable
two-body densities using the Levy-constrained search~\cite{Levy79}.
The energy (and all other observables) can be written as a functional
of $n^{(2)}$ as
\begin{equation}
E[n^{(2)}] = T[n^{(2)}] + \int d{\bf r} d{\bf r}' w({\bf r},{\bf r'})
n^{(2)}({\bf r},{\bf r'}).
\label{eqn:toten}
\end{equation}
$T[n^{(2)}]$ is a universal functional of $n^{(2)}$, that is, the
kinetic energy can be determined by knowing only $n^{(2)}$.  Whether
the system is composed of bosons or fermions enters only in the form
of $T[n^{(2)}]$; the functional dependence of the potential energy
term on $n^{(2)}$ is the same for bosons or fermions.

In the original DFT of HK, the universal functional includes the
kinetic energy and the {\it pair interaction} and is a function of the
one-body density ($F_{\mbox{\scriptsize{HK}}}[n]$).  The analogue of
$F_{\mbox{\scriptsize{HK}}}[n]$ in pair-density functional theory is
simply the kinetic energy $T[n^{(2)}]$, i.e. it does not include {\it
any} of the potential energy terms.
\begin{figure}
\setlength\fboxsep{0pt} \centering
\scalebox{0.70}{\includegraphics[angle=-90]{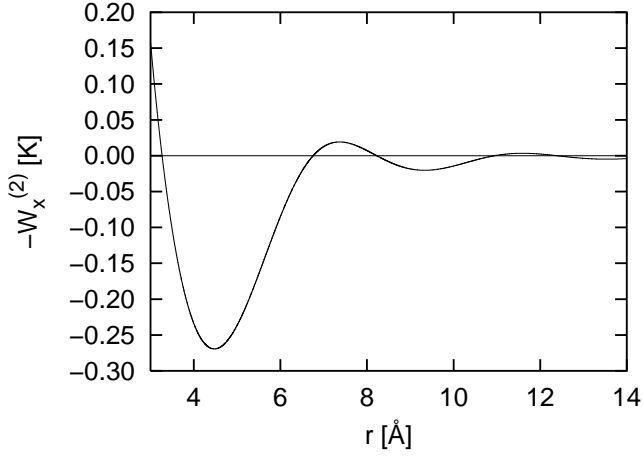}}
\caption{Effective exchange pair potential.}
\label{fig:wx}
\end{figure}

In order to obtain an applicable algorithm, we introduce an auxiliary
system of interacting bosons, and add the required corrections.  The
issue of representability of a fermionic $n^{(2)}$ by a bosonic one
shall be addressed in our extended study.  Our starting equation is
then
\begin{equation}
E_{\mbox{\scriptsize{F}}}[n^{(2)}] =
T_{\mbox{\scriptsize{B}}}[n^{(2)}] + \Delta T[n^{(2)}]+ \int d{\bf r}
d{\bf r'} w({\bf r},{\bf r'}) n^{(2)}({\bf r},{\bf r'}),
\label{eqn:fermen}
\end{equation}
where $T_{\mbox{\scriptsize{B}}}[n^{(2)}]$ is the kinetic energy of a
system of bosons with two-body density $n^{(2)}$, and
\begin{equation}
\Delta T[n^{(2)}] = T_{\mbox{\scriptsize{F}}}[n^{(2)}] -
T_{\mbox{\scriptsize{B}}}[n^{(2)}],
\label{eqn:deltaT}
\end{equation}
where $T_{\mbox{\scriptsize{F}}}[n^{(2)}]$ is the kinetic energy of a
system of fermions with two-body density $n^{(2)}$.

The correction term in Eq. (\ref{eqn:deltaT}) is the difference
between the kinetic energies of two systems identical $n^{(2)}$'s.
differing only in that one is a system of fermions, the other a system
of bosons.  We develop an approximation to $\Delta T$ by constructing
two trial wave-functions (one fermionic, one bosonic) with
approximately equal $n^{(2)}$'s, and taking the difference of the
kinetic energy expressions.  The approximation presented here is valid
for homogeneous systems.

For the fermionic one we take a wavefunction of the Jastrow-Slater
form
\begin{equation}
\Psi_{\mbox{\scriptsize{F}}} = D^{\uparrow} D^{\downarrow} F
\label{eqn:wff},
\end{equation}
where
\begin{equation}
F = \prod_{i<j} f(r_{ij}),
\end{equation}
and where $D^{\uparrow}$($D^{\downarrow}$) indicates a Slater
determinant of plane waves between atoms of spin up(down), and $f(r)$
is a correlation factor.  In constructing a bosonic wavefunction with
the same two-body density we take the same correlation factor as in
Eq. (\ref{eqn:wff}), but to account for the determinants we multiply
by additional correlation factors between parallel spins, i.e.
\begin{equation}
\Psi_{\mbox{\scriptsize{B}}} = F^{\uparrow}_x F^{\downarrow}_x F
\label{eqn:wfb},
\end{equation}
where
\begin{equation}
F^{\uparrow(\downarrow)}_x = \prod_{i<j} f^{\uparrow(\downarrow)}_x(r_{ij}),
\label{eqn:wfbx}
\end{equation}
where the product in Eq. (\ref{eqn:wfbx}) indicates a multiplication
between pairs of parallel spins.  The correlation factors
$f^{\uparrow(\downarrow)}_x(r)$ should be chosen in such a way that
the two-body densities obtained from Eqs. (\ref{eqn:wff}) and
(\ref{eqn:wfb}) are identical.  The correction term in this case can
be explicitly obtained
\begin{eqnarray}
\Delta T [n^{(2)}]  = T_0 
+ \sum_i\frac{1}{2mN_{\mbox{\scriptsize{F}}}}
\int d{\bf R} | D^{\uparrow} D^{\downarrow} |^2 
\nabla_i F \cdot \nabla_i F \nonumber \\ 
- \sum_i \frac{1}{2mN_{\mbox{\scriptsize{B}}}}
\int d{\bf R} |F^{\uparrow}_x F^{\downarrow}_x |^2 
\nabla_i F \cdot \nabla_i F \nonumber \\ 
+ \sum_i \frac{1}{2mN_{\mbox{\scriptsize{F}}}} 
\int d {\bf R} F^2 (F^{\uparrow}_x F^{\downarrow}_x 
\nabla_i^2 F^{\uparrow}_x F^{\downarrow}_x),
\label{eqn:deltaT_approx}
\end{eqnarray}
where ${\bf R}$ denotes all coordinates, $m$ denotes the mass, and
$N_{\mbox{\scriptsize{F}}},N_{\mbox{\scriptsize{B}}}$ are
normalization integrals.  $T_0 = 3/5
k_{\mbox{\scriptsize{F}}}^2/(2m)$, which is the kinetic energy of the
homogeneous non-interacting system ($k_{\mbox{\scriptsize{F}}}$ is the
Fermi wave vector).  If the correlation factor $f_x$ is chosen such
that the two-body and three-body densities obtained from the
determinants in Eq. (\ref{eqn:wff}) are identical to those obtained
from the correlation factors $f_x$ in Eq. (\ref{eqn:wfb}) then the
second and third terms in Eq. (\ref{eqn:deltaT_approx}) cancel
resulting in
\begin{equation}
\Delta T [n^{(2)}] = T_0 + \sum_i \frac{1}{2m N_{\mbox{\scriptsize{F}}}} 
\int d {\bf R} F^2 (F^{\uparrow}_x F^{\downarrow}_x 
\nabla_i^2 F^{\uparrow}_x F^{\downarrow}_x),
\label{eqn:deltaT_approxx}
\end{equation}

In obtaining a first approximation to the correlation term
$f^{\uparrow(\downarrow)}_x$ in the case of a homogeneous system we
can make use of the radial distribution function of the noninteracting
fermion gas, given by
\begin{equation}
g_x(r) = 1 - \left\{
\frac{3}{k_{\mbox{\scriptsize{F}}}^3 r^3} 
( 
\mbox{sin} k_{\mbox{\scriptsize{F}}} r 
-
k_{\mbox{\scriptsize{F}}} r \mbox{cos} k_{\mbox{\scriptsize{F}}} r
)
\right\}^2
\end{equation}
We obtain $f_x$ from an inverse hypernetted chain equation (we also
tried using $f_x=g_x$, and obtained very similar results).  While
$f^{\uparrow(\downarrow)}_x$ does not guarantee that the second and
third terms in Eq. (\ref{eqn:deltaT_approx}) will cancel, in this work
we assume Eq. (\ref{eqn:deltaT_approxx}) as the form for our
approximation.  Thus the correction term resulting from
Eq. (\ref{eqn:deltaT_approxx}) in our scheme is the sum of a two-body
and a three-body interaction,
\begin{eqnarray}
w^{(2)}_x (r) = \frac{1}{2m} \left\{ \frac{\partial^2
\mbox{ln}f_x(r)}{ \partial r^2} + \frac{2}{r} \frac{ \partial
\mbox{ln}f_x(r)}{ \partial r} \right\} \nonumber \\ 
w^{(3)}_x(r_{12},r_{13}) = \frac{1}{2m} ( \nabla_1 \mbox{ln} f_x(r_{12}) \cdot
\nabla_1 \mbox{ln} f_x(r_{13})),
\label{eqn:wx}
\end{eqnarray}
and a constant term $T_0$.  The expressions are the same for both
$f_x^{\uparrow}$, $f_x^{\downarrow}$, however in general
$f_x^{\uparrow} \neq f_x^{\downarrow}$.  Since in this work we will
deal with the unpolarized case, where $f_x^{\uparrow} =
f_x^{\downarrow}$ from now on we will drop the arrows.
\begin{figure}
\setlength\fboxsep{0pt} \centering
\scalebox{0.40}{\includegraphics[angle=0]{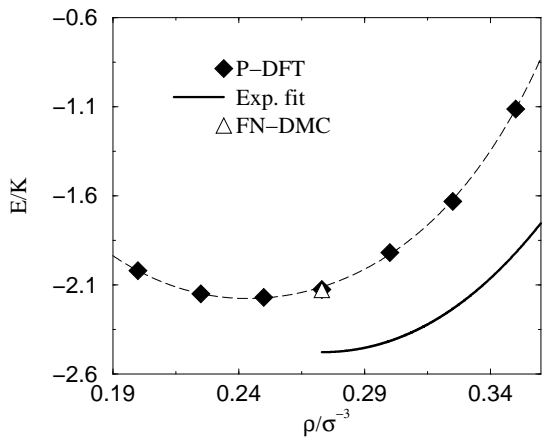}}
\caption{Total energy for $^3$He as a function of density.  The solid
curve is a fit to experimental results (Ref. \onlinecite{Aziz73}), the
curve connecting our calculated points is a fifth order polynomial
fit.  The triangle is the result from a fixed node DMC calculation
without backflow (Ref.\onlinecite{Casulleras00}).  }
\label{fig:en}
\end{figure}

Going beyond the weak coupling approximation would lead to correction
terms including the correlation factor $f$.  In principle $f$ is a
functional of the two-body density of the system, thus a
self-consistent algorithm would be necessary.  Possibly this can be
avoided by using known correlation factors for a given system under
investigation (or obtaining one from solving the Euler
equation~\cite{Moroni95}).  Potentially, better approximations can
also be obtained for $\Delta T$ if better wavefunctions are chosen in
Eqs. (\ref{eqn:wff}) and (\ref{eqn:wfb}).  In our scheme three-body
correlations and Feynman-Cohen back-flow~\cite{Feynman56} have not
been considered.

As a test of the quality of $f_x$ we have performed a Monte Carlo
simulation at the experimental density ($\rho_0 = 0.273
N/\sigma^{-3}$), and have found that the radial distribution function
obtained is in good agreement with that of the noninteracting Fermi
gas.  Obviously, in this case the energies between the fermionic
system and the auxiliary bosonic one correspond.  In Fig. \ref{fig:wx}
the effective pair potential that incorporates the nodal surface
(-$w^{(2)}_x$) is shown at $\rho_0$.  The potential is an effective
way of including the nodal structure, it is repulsive at short
distances, and displays alternating valleys and barriers of decreasing
magnitude.

We applied the above procedure to a system of unpolarized $^3$He atoms
at zero temperature.  We have used the HFD-HE2 interaction potential
due to Aziz {\it et al}~\cite{Aziz79}.  Between particles with
parallel spin the potential interaction modified by the additive terms
given in Eq. (\ref{eqn:deltaT_approxx}).  To reduce the variance in DMC
we have used a guiding function of the form given in
Eq. (\ref{eqn:wfb}) with $f(r) = \mbox{exp} -b^5/2r^5$
($b=1.15\sigma$, where $\sigma=2.556$ \AA).  The parameter $b$ was
optimized by a variational Monte Carlo calculation.  For other
calculations on the same system see
Refs. \onlinecite{Lee81,Schmidt81,Panoff89,Moroni95,Casulleras00}.

We perform a series of calculations using the standard bosonic
DMC\cite{Anderson75} algorithm.  Our cell included 108 particles in
all cases, we used an imaginary time step of 50 a.u.  We collected
averages over 100,000 steps.  Total energies are estimated in the
standard way, coordinate dependent observables were estimated using
pure estimators~\cite{Casulleras95}.  The non-coordinate dependent
part of the fermionic kinetic energy was obtained by subtracting from
the total energy the potential and the correction term in
Eq. (\ref{eqn:wx}).
\begin{table}
\begin{center}
%\begin{tabular*}{70mm}{@{\extracolsep{\fill}}r|rrr}
\begin{tabular}{|c|c|c|c|c|}\hline
$\rho/N\sigma^{-3}$ & E & V & T & $\Delta$ T - T$_0$ \\ \hline \hline
$0.2$   & $-2.050(5)$ & $-9.83(4)$  & $7.78(4)$  & $-1.040(4)$ \\ \hline
$0.225$ & $-2.174(6)$ & $-11.25(4)$ & $9.08(4)$  & $-1.143(5)$ \\ \hline
$0.25$  & $-2.220(6)$ & $-12.73(3)$ & $10.51(4)$ & $-1.253(1)$ \\ \hline
$0.273$ & $-2.147(7)$ & $-14.03(3)$ & $11.89(4)$ & $-1.341(3)$ \\ \hline
$0.3$   & $-1.944(8)$ & $-15.67(4)$ & $13.73(4)$ & $-1.521(4)$ \\ \hline
$0.325$ & $-1.65(1)$  & $-17.09(7)$ & $15.44(8)$ & $-1.521(4)$ \\ \hline
$0.35$  & $-1.17(1)$  & $-18.62(5)$ & $17.45(6)$ & $-1.642(5)$ \\ \hline
\end{tabular}
\end{center}
\caption[]{ Energy quantities per particle as a function of
density. All data are in Kelvin.}
\label{tab:en}
\end{table}

\begin{figure}
\setlength\fboxsep{0pt} \centering
\scalebox{0.70}{\includegraphics[angle=-90]{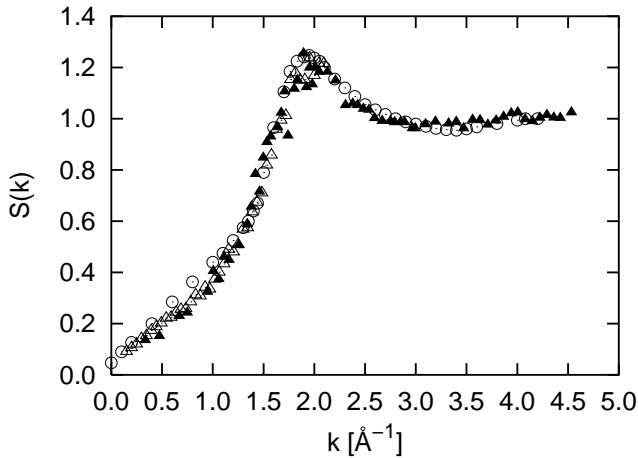}}
\caption{Comparison of calculated (solid triangles) and experimental
(circles: Ref. \onlinecite{Achter69} and empty triangles:
Ref. \onlinecite{Hallock72}) structure factor of $^3$He at
equilibrium density.  }
\label{fig:sk}
\end{figure}

In Fig. \ref{fig:en} we compare our calculated total energies with
experimental results.  The thick solid curve is a fit to experimental
results~\cite{Aziz73}.  The minimum density obtained by us is in close
agreement with the experimental result
(exp:$0.273N\sigma^{-3}$,calc:$0.244N\sigma^{-3}$). The structure
factor, shown in Fig. \ref{fig:sk} also compares well with
experiment. The experimental~\cite{DeBruynOubuter87} minimum energy is
$-2.473$ K, our calculated energy at that density is $-2.147(7)$ K.
Our energy at the calculated minimum density (from the fitted curve)
is \mbox{$-2.220(6)$ K}.  It is important to note that our results are
in good agreement with previous calculations that do not include the
back-flow correction (in Ref. \onlinecite{Casulleras00} a QMC
calculation without back-flow is reported resulting in $-2.128(15)$ K
for the total energy per particle).  We are not aware of fixed-node
DMC calculations without back-flow for the full density range
presented here, but variational Monte Carlo~\cite{Schmidt81,Moroni95}
calculations indicate that the nedlect of back-flow corrections leads
to an over-estimation of the energy by a few tenths of a Kelvin (see
Figure 1. in Ref. \onlinecite{Schmidt81}).  Our calculated energies
differ more from the experimental ones at higher densities.  This is
not surprising, since it is known that the back-flow approximation is
more crucial at higher densities.~\cite{Kwon98}

In Table \ref{tab:en} we present values for the total, potential,
kinetic energies, and for the coordinate dependent part of the
correction term ($\Delta T - T_0$).  Our value for the potential
energy at $\rho_0$ ($-14.03(3)$ K) also compares reasonably well with
other theoretical results ($\langle V \rangle = -14.84(10)$ K
\cite{Panoff89}.  The coordinate dependent part of $\Delta T$ gives
only a small correction compared to the value of the kinetic energy
itself (the bosonic kinetic energy plus $T_0$ is already a reasonable
approximation to the kinetic energy).  Thus using an auxiliary bosonic
system is a promising scheme for developing approximations.

We have demonstrated that using an algorithm constructed from the
pair-density a good description of an interacting fermionic system can
be obtained.  Our algorithm is arrived at by invoking an auxiliary
system of bosons, therefore the calculation itself can be performed by
a bosonic DMC algorithm.  While it is clear that further work needs to
be done to obtain a better approximation, the fact that we have
obtained quantitative results for the observables calculated
demonstrates that this avenue is worth pursuing.  Our future
directions include the developing and testing of more sophisticated
approximations for the kinetic energy correction term used here, such
as implementing back-flow, three-body correlation, and inverting the
fermionic hypernetted-chain
approximation~\cite{Fantoni74,Krotscheck71}.  Comparison of our method
to related ones~\cite{Reynolds82,Ceperley80,Baer98} is also of
interest.

This research was supported by MIUR-2001/025/498 and by SISSA.  We
benefitted greatly from discussions with Professor K. E. Schmidt.

%%%%%%%%%%%%%%%%%%%%%%%%%%%%%%%%%%%%%%%%%%%%%%%%%%%%%%%%%%%%%%%%%%%%%%%
%                        TABLES                                       %
%%%%%%%%%%%%%%%%%%%%%%%%%%%%%%%%%%%%%%%%%%%%%%%%%%%%%%%%%%%%%%%%%%%%%%%

\end{document}